\PassOptionsToPackage{table,xcdraw}{xcolor}
\documentclass[sigconf, authorversion, screen]{acmart}

\pdfoutput=1

\AtBeginDocument{%
  \providecommand\BibTeX{{%
    \normalfont B\kern-0.5em{\scshape i\kern-0.25em b}\kern-0.8em\TeX}}}

\copyrightyear{2021} 
\acmYear{2021} 
\setcopyright{acmlicensed}\acmConference[FDG'21]{The 16th International Conference on the Foundations of Digital Games (FDG) 2021}{August 3--6, 2021}{Montreal, QC, Canada}
\acmBooktitle{The 16th International Conference on the Foundations of Digital Games (FDG) 2021 (FDG'21), August 3--6, 2021, Montreal, QC, Canada}
\acmPrice{15.00}
\acmDOI{10.1145/3472538.3472573}
\acmISBN{978-1-4503-8422-3/21/08}

\begin{document}

\title{Design-Driven Requirements for Computationally Co-Creative Game AI Design Tools}

\author{Nathan Partlan}
\email{partlan.n@northeastern.edu}
\orcid{0000-0002-9453-2117}
\affiliation{%
  \institution{Northeastern University}
  \streetaddress{360 Huntington Avenue}
  \city{Boston}
  \state{Massachusetts}
  \country{United States}
  \postcode{02115}
}

\author{Erica Kleinman}
\email{emkleinm@ucsc.edu}
\affiliation{%
  \institution{University of California, Santa Cruz}
  \streetaddress{1156 High St.}
  \city{Santa Cruz}
  \state{CA}
  \country{United States}
  \postcode{95064}
}

\author{Jim Howe}
\email{howe.ha@northeastern.edu}
\affiliation{%
  \institution{Northeastern University}
  \streetaddress{360 Huntington Avenue}
  \city{Boston}
  \state{Massachusetts}
  \country{United States}
  \postcode{02115}
}

\author{Sabbir Ahmad}
\email{ahmad.sab@northeastern.edu}
\affiliation{%
  \institution{Northeastern University}
  \streetaddress{360 Huntington Avenue}
  \city{Boston}
  \state{Massachusetts}
  \country{United States}
  \postcode{02115}
}

\author{Stacy Marsella}
\email{s.marsella@northeastern.edu}
\orcid{0000-0002-5711-7934}
\affiliation{%
  \institution{Northeastern University}
  \streetaddress{360 Huntington Avenue}
  \city{Boston}
  \state{Massachusetts}
  \country{United States}
  \postcode{02115}
}

\author{Magy Seif El-Nasr}
\email{mseifeln@ucsc.edu}
\orcid{0000-0002-7808-1686}
\affiliation{%
  \institution{University of California, Santa Cruz}
  \streetaddress{1156 High St.}
  \city{Santa Cruz}
  \state{CA}
  \country{United States}
  \postcode{95064}
}

\renewcommand{\shortauthors}{Partlan et al.}

\begin{abstract}
Game AI designers must manage complex interactions between the AI character, the game world, and the player, while achieving their design visions. Computational co-creativity tools can aid them, but first, AI and HCI researchers must gather requirements and determine design heuristics to build effective co-creative tools. In this work, we present a participatory design study that categorizes and analyzes game AI designers' workflows, goals, and expectations for such tools. We evince deep connections between game AI design and the design of co-creative tools, and present implications for future co-creativity tool research and development.
\end{abstract}



\maketitle

\section{Introduction}

Design of AI behavior for games presents major challenges for designers, HCI and AI researchers alike. For game designers, the challenges lie in managing the complex interactions between the AI character, the game world, and the player. They must solve these while balancing the competing goals of creating compelling experiences for players, maintaining performant and stable behavior, and working with arcane languages and architectures. Designers with fewer resources to hire AI engineers may be forced to face these challenges alone or without sufficient support.

One challenge for AI and HCI researchers, then, is to aid designers in building compelling game AI. Computational co-creativity tools, in which AI assists and collaborates with the designer~\cite{davis_enactive_2015}, present a compelling option for aiding designers through these challenges, and for enabling them to build more complex, more capable, and more robust AI characters that fit their vision. If sufficiently helpful, these tools might provide new ideas, support designers' goals, and enable learning and inspection of AI architectures. However, in game design, co-creative tools have primarily been focused on map design~\cite{smith_tanagra:_2010,liapis_sentient_2013,baldwin_towards_2017,alvarez_fostering_2018,guzdial_friend_2019,alvarez_learning_2020}, rather than AI and behavior design. Craveirinha and Roque investigated co-creatively generating or modifying game mechanics~\cite{craveirinha_exploring_2016}, but their tool does not create character behavior within existing rules. Guzdial and Reidl~\cite{guzdial_interaction_2019} note that in co-creativity, ``the question of how to design the interaction between ML algorithm and human creator remains under-explored.''

To build effective tools, we need to know how AI designers prefer to work, and what they need from their tools. Thus, the first step towards building co-creative tools for game AI design is to gather requirements from AI designers. In this study, we present a participatory design-based~\cite{elizarova_participatory_2017} exploration of game AI design, asking two experienced designers about their current workflows and prompting them to imagine future tools and workflows for co-creation with AI-enabled tools. In a recent survey of creativity support tools (CSTs), Frich et al.~\cite{frich_mapping_2019} note ``an obvious, untapped potential for HCI research on CSTs to extend into studies of more complex systems employed by expert users.'' By including expert designers from the beginning of the process, we aim to ensure that the resulting tools will meet their needs. While this study focuses on game design, similar approaches may begin to answer the broader need, as noted by Abdul et al.~\cite{abdul_trends_2018}, for combined HCI and AI research towards transparent, reliable, and human-focused tools.

Recent work by Lai, Latham, and Leymarie~\cite{lai_towards_2020} begins to bridge the gap between mixed-initiative PCG tools and practical game design. They propose three ``pillars'' for mixed-initiative interfaces, based on their practical game design experience: ``respect designer control,'' ``respect the creative process,'' and ``respect existing work processes.'' Our findings include requirements that encompass each of these suggestions, as discussed below. We also go beyond these pillars, providing recommendations for interpretability, learnability, and anticipation of needs. Additionally, we explore the detailed requirements for AI design tools, specifically, and how they relate to gameplay. Zhu et al.~\cite{zhu_player-ai_2021} explore connections between gameplay and AI-human interaction, arguing that ``AI as play can expand current notions of human-AI interaction, which are predominantly productivity-based.'' They suggest using play to help people discover capabilities of an AI-enabled tool. Our work comports with these suggestions and reveals that game designers already see deep connections between their workflows and the playful experiences they are creating.

This study provides a requirements analysis and discussion of the opportunities, challenges, and potential paradigms for co-creative game AI design tools. As our primary contribution, we detail 13 categories, each with 2-3 subcategories, to organize these requirements. These categories are: \emph{``Reliable and efficient iteration''}, \emph{``Better tooling presents challenges''}, \emph{``Understandable and learnable''}, \emph{``Designers direct implementation as editors''}, \emph{``Designer has full control''}, \emph{``Anticipate and ask about needs''}, \emph{``Enable standard tasks and workflows''}, \emph{``Empower game-specific AI behavior''}, \emph{``Create varied, characterful behavior''}, \emph{``Help find and fix specific problems''}, \emph{``Communicate reasoning to the designer''}, \emph{``Communicate reasoning to the player''}, and \emph{``Player experience determines success.''} We analyze and evince deep connections between game AI design and co-creative tool design. Next, we explore relationships to prior work in game design and tool design heuristics and models, showing how our results support, expand, and complicate that work. Finally, we present implications and opportunities for future work in HCI and AI for computationally co-creative tools for design.

\section{Methods}

\subsection{Participatory Design Study}

This study represents one stage of an effort to build a computationally co-creative tool for game AI designers. We chose to use participatory design to better include the perspectives and meet the needs of this intended audience \cite{elizarova_participatory_2017,hsieh_8_2018,mozilla_participatory_nodate}. To situate the work in the context of a specific design problem for the participants, who would not necessarily be experts in AI and AI-enabled tools, we provided some priming information about behavior trees in the first session, and AI approaches we were considering for the tool in the second session, and we encouraged them to choose a specific game genre as a focus. Though participatory design is usually focused on a particular tool and continues throughout the full design process, we present this stage separately, as many of the resulting requirements are relevant to broader co-creative design tool contexts. We discuss how the priming may limit this generalization of the results in section \ref{sec:limitations}.

\subsubsection{Recruitment and Introduction}

We recruited two experienced game AI designers by individual e-mail. To be eligible for the study, we required participants to be over 18 years of age, to be able to read and write in English, and to have some experience designing AI for games (either hobbyist or professional). We provided no compensation.

Due to the COVID-19 pandemic, we ran the study as a series of three virtual one-hour group participatory design sessions, followed by an individual one-hour virtual interview. We received informed consent from the participants over encrypted video conferencing software, and we also received permission to record audio and screens, and to use and save information from virtual whiteboard and collaborative document editing software. After the study sessions were complete, we transcribed the recordings and performed the analysis on the transcripts.

We collected demographic data on the participants' experience with game AI design, and specifically with behavior trees and evolutionary algorithms. We did not collect age, gender, or race, as we do not expect those to have a significant effect on the results not already captured by design experience, and because we would not have sufficient participant variety, in any case, to make meaningful inferences from those variables. We discuss the implications of the participant demographics in Limitations, below.

One researcher was responsible for all communication with the participants, asking and answering questions to help guide the discussion when necessary. We attempted to keep such guidance limited, with the goal of primarily adding new questions and topics, rather than limiting creativity or discussion. Each participatory design session was structured as a conversation among the two designers and the researcher.

\subsubsection{Session 1: Current Process and Tools for Game AI Design}

In the first session, after the informed consent and demographic survey, we began by framing the session goals: for participants to consider and describe their current tools, tasks, and methods for designing game AI, to provide context and ground future designs for a new AI-enabled tool. To focus this discussion, we asked the participants to collaboratively outline a fictional rendition of their typical workflow, in the context of their current or previously-used tools, focusing on a particular type of game of their choice.

We framed the discussion with the participants around \emph{behavior trees}, a commonly-used behavior definition structure and paradigm for game AI \cite{isla_handling_2005}. The designers also discussed \emph{hierarchical finite state machines (HFSMs)}, another common system for game AI \cite{girault_hierarchical_1999}. These form a common ground for discussing the designers' prior experience and current workflows. We asked follow-up questions related to their discussion, about collaboration with other team members, use of the aforementioned AI technologies to structure the behavior, and testing and debugging.

\subsubsection{Session 2: Imagining a Better Workflow}

In the second participatory design session, we began with a brief recap of the previous session. Next, we explained the purpose of this session: to imagine a more desirable and effective workflow, enabled by a computationally co-creative AI. To frame the discussion, we introduced genetic algorithms, which can modify and act on existing artifacts, including tree structures~\cite{koza_hierarchical_1989}, improving them towards a particular objective. Genetic algorithms have been used with behavior trees~\cite{lim_evolving_2010}. We also introduced computational co-creativity~\cite{davis_enactive_2015}, and specifically mixed-initiative interaction paradigms~\cite{liapis_mixed-initiative_2016}, to frame how game designers might work with an AI.

Then, we asked the participants to explain how they would imagine working with such a tool, adding follow-up questions to elicit additional information. These included asking how they would want to specify goals and success for the AI behavior, whether and how they would act as proxies for players in testing that behavior, and how they would want to visualize and evaluate possible alternatives for the resulting behavior.

\subsubsection{Session 3: Interface Design}

In the third and final session, we reviewed the previous session and then asked the participants to devise a concrete interface design for the workflow they had imagined, with the use of diagrams and writing as necessary. We encouraged the participants to iterate on and flesh out the ideas from their prior discussions as necessary. As in previous sessions, we asked follow-up questions to continue eliciting design details, such as how they would imagine the interface enabling them to define behavior, and how they would want to see and evaluate the results.

\subsubsection{Post-Interviews}

In the post-interview, held individually with each participant and a single researcher, we asked reflection questions about the study and the design discussions that occurred. We prepared some questions in advance, but also asked follow-up questions to expand on relevant details. The guiding questions included:

\begin{itemize}
\item How was your design inspired by existing tools, and how does it differ from/improve on them?
\item What parts of the interface you designed are you most happy with/presented challenges, and why?
\item What part of the interface do you think will be intuitive, and what parts do you think users will struggle with?
\item Would you use the interface that you have designed? If so, for what purposes? If not, why not?
\end{itemize}

\subsection{Analysis Protocol}
\label{analysismethods}

We performed qualitative data analysis on our notes, transcripts, and the participants' written materials using the following thematic analysis protocol.

\subsubsection{First-Cycle Coding: In-Vivo, Process, and Values}

In the initial coding process, our goal was to build familiarity with and insight into the data by adding a rough but comprehensive set of initial annotations (codes), along with accompanying notes about our thought process (memos). One researcher individually coded the data, starting with each session transcript in chronological order, then returning to code the various additional researcher notes and participant writings. We simultaneously employed the following three methods, which are further detailed by Saldaña~\cite{saldana_coding_2015}.

\begin{enumerate}
\item \emph{In-Vivo Coding}, where codes are particularly important short phrases directly from the participants' own language. These should encapsulate a larger section of information, or highlight language specific to the context/participant.

\item \emph{Process Coding}, using ``-ing'' verbs (gerunds) to indicate processes or actions participants are taking or describing.

\item \emph{Values Coding}, in which we code for concepts that participants place positive or negative value on. These may be ethical values, but they may also be utilitarian values (e.g., this is helpful or useful to me), or other types of value.
\end{enumerate}

As the coder derived new codes, they wrote analytic memos containing a brief description of the definition of that code and how it should be used, as well as their thoughts on possible relationships with other codes.

\subsubsection{Post First-Cycle Analysis: Code Mapping}

After completing the first cycle coding, we performed \emph{Code Mapping}, as described by Saldaña~\cite{saldana_coding_2015}. In this process, we extracted the codes into a list. Then, we grouped the codes into related groups, and wrote category headers that might apply to those groups. Where first-cycle codes were highly related and, in our opinion, could be consolidated for brevity and clarity, we additionally defined potential sub-categories that combined several of the initial first-cycle codes into more concise form. Three researchers met to discuss these.

\subsubsection{Second-Cycle Coding: Focused Coding}

In the second-cycle coding process, the same researcher who performed the first-cycle coding re-coded the data. We used \emph{Focused Coding}~\cite{saldana_coding_2015}, in which the first-cycle codes are grouped into loose categories, which are themselves termed as codes, developed through the Code Mapping process described above. Finally, the first researcher selected a representative subset of the data for a second researcher to separately code. We used this subset to compute inter-rater reliability (IRR) using Cohen's Kappa~\cite{cohen_coefficient_1960}. We found an initial result of 0.65, and we met to adjust for code overlaps and unclear definitions. After adjusting the code book, we conducted a second round of IRR, resulting in an agreement of 0.7, indicating strong agreement~\cite{landis_measurement_1977}.

\subsubsection{Final Analysis: Category Descriptions, Member Checking}

After second-cycle coding was complete, we met to expand the codes that had emerged from the second-cycle coding into detailed categories, fully defined with descriptions and sub-categories. Derived from the second-cycle codes, which were derived from the first-cycle codes, the categories gather and connect design insights and requirements across the sessions. We performed member checking by presenting the final analysis -- the categories, their descriptions, and subcategories -- to the participants for comment.

\section{Results}

\subsection{Participant Demographics}

We recruited two designers through direct e-mail recruitment, designated Blue and Gold. Both have 11 or more years of experience with professional game development and/or research, with multiple years of independent or educational experience on top of that. Blue has worked on 6-10 released games, while Gold has worked on 1-5. Blue's experience included 1-5 years specifically working with game AI, while Gold's included 6-10 years. While both designers knew something about genetic algorithms, both reported their experience level at only 1 out of a possible 5. With behavior trees, Gold ranked their experience as 4 of 5, while Blue ranked their experience as 2. Neither designer had any experience with evolving behavior trees. Blue had participated in a pilot study related to this work, in which we asked them to simulate performing several tasks on a non-functional mockup user interface design for a tool for working with behavior trees using evolutionary algorithms. Thus, they had some prior understanding of how evolutionary algorithms might interact with behavior trees for game AI. We described the concepts of evolutionary algorithms and computationally co-creative tools to both participants before asking them to begin designing a tool to work with those concepts in Session 2, to provide similar theoretical understanding to both.

\subsection{Primary Topics and Connections}

\subsubsection{Session 1: Current Workflows, Tools, and Player-Focused AI Design}

The first session centered around describing the designers' current workflows and considerations for creating AI behavior. We first prompted them to decide on a scenario to focus their discussions. They chose to imagine a stealth game, in which a ``swarm of finder robots'' would team up to search for the player. Blue explained that this would require complex AI reasoning: ``having AIs that need to choose what their decisions are based on whether the player currently perceives them or not adds a whole lot of layers of things that make AI really interesting.'' In this vein, Blue described how crafting a particular player experience would drive their AI design: ``it would be important that... behaviors also have flaws that are not removed entirely, they need to be ultimately exploitable by the player to make them feel smart and have fun.'' To ensure this comprehensibility of AI actions, Gold said, ``with a stealth game you need to give feedback [...] about, you know, their level of awareness.'' Throughout, the discussion emphasized player-focused considerations, such as clear feedback and a feeling of fairness.

The designers continued by discussing the specific elements that compose game AI, in stealth games but also in general, such as perception, memory, states and transition logic, and communication with other game systems, such as animation. Gold explicitly tied control over AI behavior to the resulting player experience: ``once you’ve gone into a disguise or hiding somewhere, and they’re, like, checking spots where you might be, I like actually having my finger on the – on the button there as a designer [...] if this happens three times in a row, like, don’t let them find you. Like, that’s just a bad experience.'' The designers connected this need for control to create their desired player experience with a need to understand and visualize the AI behavior for themselves. Gold said, ``So this is getting a little bit into the analytics around these AIs as well. I would definitely want to understand, you know, after the fact, what decisions they made and why, and have some AI... some metrics kind of recording that so that I can tune it.''

Next, the participants explained how they work with a team, and with existing tools: in most cases, they provide specifications to an engineer for the initial implementation, and later iteratively tweak it. Gold said of the behavior, ``I’m more, you know, of an editor of that and high-level designer of it.'' They added that it would help to have data-driven means of changing parameters. Gold said, ``I always like to do, you know, tuning of things in a data-driven way as much as possible [...] Be able to see, you know, a lot of the variables that are gonna be driving these things in like a spreadsheet format, and be able to in the game also see a lot of visual feedback on things.'' Finally, they explained how they would debug and gather playtesting feedback. They described a desire for visual debugging features, and for replay of particular situations. Blue said, ``So, imagine you’re playing through it and you see a... an agent do something, or a set of agents do something interesting, and you’re like, `Oh man, how did that happen?' Uh... being able to rewind back to it and snapshotting that so you can share that – that state and time, where the agents were, the positions in the game,'' adding that ``it would be awesome if you could do that, but it always seems like a thing that’s probably super difficult.''

\subsubsection{Session 2: Imagining a New Workflow Supported by a Co-Creative Tool}

In the second session, we prompted the participants to imagine a new workflow in an AI-assisted tool. They  emphasized the importance of a reliable, learnable tool that employed game AI best-practices. Gold said, ``I want one of the things that is promised to be ‘hey, we did some of the work for you, here are some basic things based off of really commonly used example, like best practices.’ So like ‘hey, we have this as a starting point, this is a, like, kind of basic stealth searching AI, here’s a basic aim and shoot behavior [...]''' Gold emphasized the need for reliable control over the results: ``I’d want some assurance that we can lock things down at some point, like ‘hey we like this behavior, alright we’ll stop trying to iterate on that’ because we want it to be predictable to players in the end.'' Here, once again, the participants tied their work requirements to player experience. Blue emphasized control as well, suggesting ``low latency iteration'' to understand the impact of changes.

While they put significant value on control over the results, the designers also suggested ways for the tool to take the initiative: it could present options, make suggestions, provide placeholders, and create surprising variety. Blue added, ``I could see myself wanting to then say `you know what, I've got the basics out here that I like,' and then I would love to hand it off to my AI collaborator and say `within this space, give me some options, play with this.''' To help designers learn, Gold said, ``it's great to be able to have a tool which asks a couple of big questions to get you started and then sets defaults. And then it doesn't present to you all the things that you can modify up front. [...] And then it can add on additional state after that, like what happens if they're blinded, what happens in this situation, and start getting out to the edge cases.'' This focus on intuitive communication and sensible defaults to aid learning persisted throughout.

As another way to aid designers, the participants suggested several ways to visualize behavior: through placeholder feedback, such as animations, UI elements, and sound; and through visualization of options, branches, and traits. Gold said, ``I’m playing with the AI first, and I’m doing an experiment and of course I’m not going to have any animation for that yet, so character turn blue in this state, character turns red. Icon appears above head. Dock a UI panel to them so that I can show icons that I made up.'' Blue also suggested pre-made scenarios for testing behavior: ``if there’s a way to say `This environment, I now want to see the seven options that you gave to me one at a time [...] Let me look at these three in more detail. [...] I've liked this one, I lock it in, and now I might do some hand tweaks.''' Gold added that it would be helpful to ``make it super clear how and where they branch and do that visually. Like `alright I branched off a drunk one at this point,' and two days later after making lots and lots of basic changes to all of the AIs, kind of see how much of an effect on the drunk one.'' Thus, both in-game and static visualization figured heavily in the discussion.

The designers again emphasized the player experience as the primary measure of success, and tied their tool design needs to it. Blue described wanting AI to be described in terms of traits and characterization, saying, ``it would almost seem magical if the tool could both generate a wide range of possibilities and then somehow summarize or characterize `I've got three cautious options you can evaluate, two aggressive options, and one drunk, whatever.' If it had ways of actually, like, looking at what I generated in a way, to help you say, like, `oh this implies a personality, or a kind of overriding mood, I guess, to the agent,' that would be very interesting, because these are the kinds of things that are appealing to players ultimately.'' Addressing success measures and goals for AI behavior, Gold emphasized: ``defining success as being sometimes failing would be important'' and explained that included ``the illusion of fairness being far more important, than any kind of mechanical success test that you can put in there. So some way to get feedback on that, for people to input that after an iteration in testing.'' Blue agreed, ``I 100\% agree that efficiently solving this is actually not what you want. Or probably not what you want from a fun for the player standpoint.''

\subsubsection{Session 3: Designing a Co-Creative Interface}

In session 3, the designers expanded on their discussion of future workflows to describe requirements for a computationally co-creative tool, concretizing the details, specific interfaces, and interactions. Gold had written a list of potential questions the tool might ask an AI designer. These included, for instance, ``How does it locomote,'' with sub-questions ``On a navmesh'' and ``Does it hover or fly,'' and then ``What behaviors does the AI have'' with several other sub-questions. In addition to reiterating and detailing the need for onboarding and expanding complexity to help designers learn to use the tool, the designers elaborated on their desires for intuitively defining traits and characterful behavior. Blue wrote a list of potential aspects of an AI character: ''Body (characteristics), Mind (behaviors), Traits (cautious, brave, distractable), Senses (perception/inputs).'' Gold agreed, ``it can’t be too technical, of, like, thinking of it in terms of, like, programming hierarchy. [...] I really like breaking it down that way, so that the player can kind of... drill down in an intuitive way to find the thing that they are thinking about.''

The designers also elaborated on ideas for visualization of AI behavior and decision-making. Gold suggested a ``[...] summary that shows what components were used after a play session, like, keeping track of how much different, you know, nodes in it were used, and what the typical paths were [...] like a heatmap [...] a worn path of like... where the tree is [...] typically running. And then see that over time, like be able to kind of compare that to other versions [...]'' Blue suggested ways to see what the agent is thinking in real time, such as ``equivalents of thought bubbles somehow, that would just tell me like, `[...] this is the current, you know, state of my mind, I’m searching.''' The designers brainstormed ways to customize these visualizations and to focus on a particular agent, to avoid overwhelming the designer.

Connecting the tool design once again to player experience, the designers further detailed how they might provide goals to and evaluate AI behavior alongside a co-creative AI tool. Gold said the tool might detect which behavior was not in line with design goals: ``it could point out, like, `Did you actually want to give this tremor-sense? It’s finding the player 90\% of the time.''' Blue suggested defining goals in terms of transitions in behavior or state: ``if somehow you were able to say, `For any state, these are the transitions that I wish for you to solve for.''' Both designers emphasized the importance of personality and character traits, sensible defaults for them, and testing playgrounds. They frequently called out places where they were uncertain, however, of the best ways to implement these features and noted potential difficulties in interface design and in providing sufficient variety to support many different types of design. They noted community and sharing among designers as potential ways to expand the available content.

\subsubsection{Reflections from the Interviews}

In the post-interviews, we asked the designers about similarities and differences from prior tools in their imagined designs. Blue noted that the HFSM architecture's clear flow of control inspired many of their suggestions. They said that the main improvements they imagined were run-time visualizations of the decision-making, and using character-like traits to specify AI behavior. Gold said that, while their ideas were related to behavior tree tools they had used, ``It doesn't feel like it was actually something that was too close to something that I've used before, it felt like it was mostly a thing that we kind of imagined up as we were going along.'' They similarly highlighted UX and UI feedback on behavior and reasoning as novel, as well as the process of asking increasingly detailed questions to help a designer create AI behavior.

We then asked the designers what parts of their designs they were confident and uncertain about, and what parts might be challenging. As potential challenges, Blue highlighted the intuitive visualization of behavior, as well as details of implementing character traits, saying, ``I think another thing that would be challenging is making sure that there’s a good way for the sort of trait encapsulations, to both be identified, and descriptive and prescriptive.'' However, Blue expressed confidence that ``agent-centered editing,'' including characterization through traits, would be useful if these challenges can be overcome. They also noted confidence in the idea of defaults and easy-to-test scenarios for rapid learning. Gold noted possible difficulties in creating and managing large numbers of options and in determining how to coordinate with the designer's goals, but expressed confidence in intuitive visualization as a helpful idea.

Each designer also expressed strong general agreement with the other about the designs. They both said they would use the resulting tool if it was effective and worked well with their goals. Blue said, ``it would need to work within the engine that the company plans to ship with. I think it could be useful as early prototyping if it’s good at getting up to speed quickly. If it’s quickly usable then you could start validating some ideas.'' In this way, the designers highlighted prototyping and ideation as a primary benefit of such a tool.

\subsection{Categorization}

\begin{table*}

\begin{tabular}{|c|l|}
    \hline
    
    \rowcolor[gray]{0.8} \textbf{Reliable and efficient iteration} & Needing confidence in tool's effectiveness and efficiency.\\
    \hline
    \multicolumn{2}{|c|}{\emph{Subcategories:} \emph{Efficient iteration}, \emph{Confidence and predictable results}} \\
     
    \hline
    
    \rowcolor[gray]{0.8} \textbf{Better tooling presents challenges} & Anticipated challenges and rewards of good tool support.\\
    \hline
    \multicolumn{2}{|c|}{\emph{Subcategories:} \emph{Expect challenges}, \emph{Value tool support}} \\
     
    \hline
    
    \rowcolor[gray]{0.8} \textbf{Understandable and learnable} & Intuitive, learnable workflow \& interface, managing complexity.\\
    \hline
    \multicolumn{2}{|c|}{\emph{Subcategories:} \emph{Intuitive and learnable}, \emph{Clear, familiar workflow}, \emph{Complexity only as necessary}} \\
     
    \hline
    
    \rowcolor[gray]{0.8} \textbf{Designers direct implementation as editors} & Designers oversee and modify results to fit their vision.\\
    \hline
    \multicolumn{2}{|c|}{\emph{Subcategories:} \emph{Designer oversees implementation}, \emph{Tool as a partner}}\\
     
    \hline
        
    \rowcolor[gray]{0.8} \textbf{Designer has full control} & Designers can inspect, edit, and finalize all details.\\
    \hline
    \multicolumn{2}{|c|}{\emph{Subcategories:} \emph{Control all details}, \emph{Changes are preserved}} \\
     
    \hline
    
    \rowcolor[gray]{0.8} \textbf{Anticipate and ask about needs} & Tool provides choices, clarifying questions, and placeholders.\\
    \hline
    \multicolumn{2}{|c|}{\emph{Subcategories:} \emph{Anticipate requirements and issues}, \emph{Propose options and ask questions}} \\
    
    \hline
    
    \rowcolor[gray]{0.8} \textbf{Enable standard tasks and workflows} & Support common tasks and patterns in game design.\\
    \hline
    \multicolumn{2}{|c|}{\emph{Subcategories:} \emph{Follow game design standards}, \emph{Standard AI tasks}} \\
    
    \hline
    
    \rowcolor[gray]{0.8} \textbf{Empower game-specific AI behavior} & Enable designers to create game-specific logic and behavior.\\
    \hline
    \multicolumn{2}{|c|}{\emph{Subcategories:} \emph{AI serves gameplay}, \emph{Connect AI into the game world}} \\
    
    \hline
    
    \rowcolor[gray]{0.8} \textbf{Create varied, characterful behavior} & Tool creates varied behavior that evokes characterization.\\
    \hline
    \multicolumn{2}{|c|}{\emph{Subcategories:} \emph{Create character}, \emph{Create variety}, \emph{Visualize tradeoffs and changes}} \\
    
    \hline
    
    \rowcolor[gray]{0.8} \textbf{Help find and fix specific problems} & Set up, test, and review specific situations to find bugs.\\
    \hline
    \multicolumn{2}{|c|}{\emph{Subcategories:} \emph{Test and review specific situations}, \emph{Solicit feedback}} \\
    
    \hline
    
    \rowcolor[gray]{0.8} \textbf{Communicate reasoning to the designer} & Visualize and have clear control flow for inspectable behavior.\\
    \hline
    \multicolumn{2}{|c|}{\emph{Subcategories:} \emph{Visualize AI reasoning and decisions}, \emph{Understandable control flow}} \\
    
    \hline
    
    \rowcolor[gray]{0.8} \textbf{Communicate reasoning to the player} & Provide player feedback, show what AI is doing and thinking.\\
    \hline
    \multicolumn{2}{|c|}{\emph{Subcategories:} \emph{Give feedback}, \emph{Give the player tools}} \\
    
    \hline
    
    \rowcolor[gray]{0.8} \textbf{Player experience determines success} & Successful AI is not perfect, but creates good player experience.\\
    \hline
    \multicolumn{2}{|c|}{\emph{Subcategories:} \emph{Measure success by player experience}, \emph{Embrace imperfection}, \emph{Maintain an illusion of intelligence}} \\
    
    \hline
    
\end{tabular}

\caption{The list of categories and subcategories produced by the analysis. Categories are listed in bold, with subcategories listed in italics below the category's brief definition. They are ordered generally from focus on designer experience towards player experience.}

\label{categorization}

\end{table*}

To comprehensively catalogue these various topics and insights, in the coding and analysis process described in section \ref{analysismethods}, we summarized and collected the requirements from the session transcripts, notes, and materials. This analysis produced 13 categories, each with 2-3 subcategories, as listed in Table \ref{categorization}. We present the details of each category's definition and attributes below, with subcategory definitions listed as bullets. We order them loosely from those primarily focused on designer experience of using the tool first, those focused on the player experience of the resulting AI later. However, they are not fully separable - every category has implications for tool design and relates to player experience.

In member checking, one participant (Gold) responded. They indicated general agreement with the results and provided two specific clarification comments, which we incorporated here: that complexity should not be fully hidden, and that the tool might educate designers as it anticipates potential issues.

\subsubsection{Reliable and efficient iteration}

This category represents the value the designers place on confidence in their tools supporting them effectively and efficiently in the design process. More specifically, the designers discuss the need for rapid iteration, with ``tactile'' and immediate feedback for each change they make.

\begin{itemize}

\item \emph{Efficient iteration}: The designers describe their process while creating the game as an iterative one. They have limited resources, and want to be assured that the tool will make each cycle of tweaking and seeing the results as immediate as possible. They describe the desire for ``tactile'' feedback and ``having a tight, low-latency loop.''

\item \emph{Confidence and predictable results}: The designers explain that they need to be confident that the tool will be effective for its specific purpose, and that it should show high-quality examples or capabilities that match its purpose. They want to feel that ``the tool has got my back.''

\end{itemize}

\subsubsection{Better tooling presents challenges}

This category represents the designers’ discussion of perceived or anticipated challenges in the design process. Often, however, they also explain the value they place on having support for these difficult tasks, or the value of tools that could overcome the challenges they describe.

\begin{itemize}

\item \emph{Expect challenges}: Designers expect several of their goals and hopes for AI-enabled tools to be difficult to implement for technical or interaction design reasons, or to require significant time to create enough content.

\item \emph{Value tool support}: Designers place a high value on supportive, well-designed tools. They have many hopes and expectations for future workflow improvements and the potential experiences those tools could help them create.
\end{itemize}

\subsubsection{Understandable and learnable}

This category represents the value the designers place on having an intuitive and clear workflow and interface for their tools. For this purpose, they want the tool to provide sensible defaults that they can use as a starting point. 

\begin{itemize}
\item \emph{Intuitive and learnable}: The designers express the need to be able to intuitively understand the language or representation in which they are building AI. They want it to use terms that they recognize or can quickly grasp, rather than terms that are arcane and engineer-focused.

\item \emph{Clear, familiar workflow}: Designers want to quickly learn a workflow, and have familiar patterns or concepts from other game design tools; they say ``there’s always a common thread'' in tool designs for AI.

\item \emph{Complexity only as necessary}: The designers explain that they would like complexity to be revealed only as it becomes necessary (though not irrevocably hidden), especially while learning. They want to be able to quickly set up high-level or basic behavior, and only later dig into the details.
\end{itemize}

\subsubsection{Designers direct implementation as editors}

This category represents the designers’ role overseeing the implementation process and how they would like to collaborate with an AI-driven tool. They describe primarily working as ``editors,'' or overseeing the process of creating AI. This category captures how the designers would take initiative to drive the creation process, whereas \emph{Anticipate and ask about needs} captures cases where the AI takes initiative.

\begin{itemize}

\item \emph{Designer oversees implementation}: Designers explain their primary role as that of an editor, providing high-level direction, oversight, and tweaks and fixes for implemented behavior. They do not always want to or know how to implement the details of their vision, and they may be willing to let a tool or team member take that role. However, they would like to provide specifications to drive the behavior, and want to review and edit the results.

\item \emph{Tool as a partner}: Designers express the desire to work with an AI-driven tool as a partner or collaborator. They hope the tool can understand their goals and work with them to make their vision a reality. They hope to push some responsibilities to the tool, and to trust it to provide relevant ideas and AI behavior as ``a collaborative, creative partner.'' They hope it will help them ``express the possibility space'' of their vision.

\end{itemize}

\subsubsection{Designer has full control}

This category represents the designers’ expectation that they will have full control over the AI behavior. They value the ability to ``drill down'' and access additional implementation details as necessary to make their desired changes, and they require data-driven, accessible methods for editing those details.

\begin{itemize}

\item \emph{Control all details}: Designers want to be able to inspect and edit any part of the AI, and every detail, when necessary. They would like to be able to ``drill down'' into each setting and parameter, and to define new concepts or settings if there is no current way to do what they want. This also means that ``data-driven'' design, where settings are quickly and easily modifiable separate from arcane code, is desirable.

\item \emph{Changes are preserved}: Designers want assurance that they will have the final word on behavior. They want to define what successful behavior looks like, and they want their changes to be preserved.

\end{itemize}

\subsubsection{Anticipate and ask about needs}

This category represents the designers’ desire for the tool to propose new ideas, by discovering potential requirements and/or solutions to problems. If the designer is taking the initiative, \emph{Designers direct implementation as editors} applies. If the tool is taking the initiative, then this category applies. The designers describe this occurring primarily through the tool providing choices, asking questions to determine requirements and goals, and providing placeholder data for the designer to specify. This may also educate the designer about new ideas.

\begin{itemize}

\item \emph{Anticipate requirements and issues}: Designers would be excited if the AI-enabled tool can determine their needs and requirements, especially if it can fill in gaps or come up with ideas they might have missed otherwise. This also extends to noticing potential issues or missing information in advance, rather than in later debugging.

\item \emph{Propose options and ask questions}: Designers would like the AI to propose various options and present alternatives. They would also like it to ask them questions to clarify uncertainties or prompt for high or low-level design information. Designers hope that the tool would be able to ask them high-level questions and use the answers to generate relevant behavior, perhaps with some placeholders for the designer to fill in afterwards. Later, they may want the tool to ask additional questions to refine the behavior further.

\end{itemize}

\subsubsection{Enable standard tasks and workflows}

This category represents the value the designers place on the tool supporting them in performing common tasks and using common patterns in game design. They describe cases where they want to use such common patterns and practices. They also describe core behavior that almost all game AI agents will need, such as memory, perception, a way of tracking state, and some separation between parts of the AI and game logic.

\begin{itemize}

\item \emph{Follow game design standards}: Designers would like to generally follow standards and ``best practices'' from other games, and they would like their tools to know about or work with those standards. This may be player-facing, such as using standard AI behaviors, or internal, such as ``separation of concerns'' between systems.

\item \emph{Standard AI tasks}: Designers want to be able to build standard AI systems that provide the basic capabilities necessary for game AI. They discuss needing AI to perceive the world, remember information, switch between various states or modes, and use logic to reason about how to behave. If there are multiple AI characters, they also need to determine if and how they will communicate or coordinate.

\end{itemize}

\subsubsection{Empower game-specific AI behavior}

This category represents the value the designers place on the tool enabling them to create game-specific logic and behavior. In this study, this primarily applies to discussions of stealth game genre-specific logic and behavior, as that was the genre chosen by the designers. This category also applies to cases where the designers are sharing information between AI systems and the rest of the game, as those are also cases where the AI interfaces with the specific game being created.

\begin{itemize}

\item \emph{AI serves gameplay}: Designers want the AI design to be informed by the needs of the gameplay. They want to build AI systems that very specifically work to achieve the particular gameplay they are creating, rarely the other way around. It is important that the tools enable the particular type of AI they need for the game, or that they can easily be modified to enable that particular AI behavior.

\item \emph{Connect AI into the game world}: The AI needs to be able to communicate with the other game systems and interact with the world. This means that the AI needs to reason about the game world, reason about space and current situations, and inter-operate or communicate with other systems. This may be enhanced with ``smart objects'' or other annotations or logic that inform AI behavior, outside of the characters themselves.

\end{itemize}

\subsubsection{Create varied, characterful behavior}

This category represents the value the designers place on being able to create varied and expressive AI characters. They hope to have tools that provide multiple options, differentiated in their behavior in characterful ways. The designers would like to create specific traits to describe aspects of these characters, and/or to have the tool provide or suggest such traits. They would like to see clear information about the tradeoffs between these characters, as well as to track changes as they or the tools modify the AI.

\begin{itemize}

\item \emph{Create character}: Designers explain how they would like their AIs to feel distinctive and cohesive in having a particular characterization. They want to ensure that each character is memorable and unique, and associated with particular traits or attributes that define them. These should be apprehensible and meaningful to the player, with traits making a significant impact on the character’s behavior.

\item \emph{Create variety}: Designers describe the need for several different AI behaviors, asking for meaningfully different, varied AI agents. They want to choose between these options, or use several of them to vary player experiences.

\item \emph{Visualize tradeoffs and changes}: Designers want the tool to show differences and tradeoffs between AI options. They want to know how characters differ, and how they were derived from previous versions or other options.

\end{itemize}

\subsubsection{Help find and fix specific problems}

This category represents the designers’ expectation that the tool should help them find and fix specific issues in their AI behavior. They hope to be able to set up and test very specific situations, with precise review and repeatable performance, and to get external playtest feedback.

\begin{itemize}

\item \emph{Test and review specific situations}: Designers would like to be able to quickly and easily set up specific situation ``playgrounds'' to test specific functionality. They would like to be able to consistently replay and review these.

\item \emph{Solicit feedback}: Designers want to show their work to other team members, or external playtesters, and receive feedback. They especially want to find bugs and ``edge cases'' that they might have missed when first implementing the AI. They may also be interested in automating that process and receiving feedback from AI tools.

\end{itemize}

\subsubsection{Communicate reasoning to the designer}

This category represents the value the designers place on having the tool communicate its choices and reasoning to them. They expect to understand how the AI character will make decisions, and clear hierarchies of control and levels of detail are significant in helping to make sense of this. They also express the desire for detailed analytics and data to inform their decisions.

\begin{itemize}

\item \emph{Visualize AI reasoning and decisions}: Designers describe the need to understand why an AI is acting in particular ways, and they propose visualizations, information, and explanations that would help them understand AI decision-making. They want to inspect the AI’s decision-making during recorded interactions.

\item \emph{Understandable control flow}: Designers want to be able to inspect the flow of control and other relevant information about the AI. Even without testing the AI in a play session, they would like to be able to reason about its behavior. This can be enabled by clear hierarchies of control and separations of state and/or logic, with clear information about how the AI will decide or change between behaviors.

\end{itemize}

\subsubsection{Communicate reasoning to the player}

This category represents the concept of designing AI to enable the player to understand its behavior. The designers explain how they use feedback to show the player what the AI is doing and thinking. They also describe making choices for behavior that will feel sensible to the player, and giving the player the tools to take advantage of that predictable behavior.

\begin{itemize}

\item \emph{Give feedback}: Designers describe the need to give clear information to the player about AI behavior. They endeavor to provide consistent feedback for player actions, and tell the player what the AI is thinking and doing.

\item \emph{Give the player tools}: Designers explain how they ensure the player can anticipate and take advantage of predictable behavior from the AI. They talk about creating consistent and intentionally exploitable behavior, and providing means for the player to note and use it to defeat AI agents.

\end{itemize}

\subsubsection{Player experience determines success}

This category represents the value that the designers place on using player experience, not ``perfect'' AI, as the primary measure of the quality and success of their AI. The designers also want to surprise and delight players by creating AI behavior that creates and maintains an illusion of intelligence.

\begin{itemize}

\item \emph{Measure success by player experience}: To determine whether the AI is successful, the designers say that the final arbiter is the player. If the player is satisfied with the AI behavior, then it is a success. Designers can act as proxies or define other goals in service of that, but ultimately, the player experience determines the success of the AI.

\item \emph{Embrace imperfection}: The designers describe using several methods to enhance player experience through imperfect behavior. For instance, they describe AI intentionally failing or acting inefficiently to let the player have a better chance of winning. They describe using ``cheating'' to enhance challenge or change the experience. They emphasize that ``perfect'' AI is not the goal, but rather achieving a particular player experience.

\item \emph{Maintain an illusion of intelligence}: Designers describe the goal of creating and maintaining the ``illusion of intelligence.'' They hope to use certain especially memorable or specific behaviors to delight the player, and they describe using callbacks or memory to make the AI appear knowledgeable. They explain the need for fallback behaviors and contingencies to maintain this illusion, rather than the AI failing to act in unexpected situations.

\end{itemize}

\section{Discussion}

\subsection{Interplay Between Game Design and Tool Design}

As detailed in the results, above, the designers frequently describe their own needs for working with an AI-enabled design tool similarly to how they explain design patterns for player-facing game AI behavior. This connection is particularly evident in the direct parallel between the categories \emph{``Communicate reasoning to the designer''} and \emph{``Communicate reasoning to the player.''} The parallel also extends to consistency and reliability of behavior, which the designers also describe as being critical both for player-facing behavior and for tool AI for game design. There is a complication here, in that designers may require additional clarity for behavioral details that are too granular to matter to players.

Designers beginning to use computationally co-creative tools are similar to players in another important sense: they are both learning an unfamiliar system and its rules. The games heuristics literature frequently highlights the need for learnability and tutorialization in games~\cite{federoff_heuristics_2002,pinelle_heuristic_2008,desurvire_game_2009}, and this study shows the need for similar tutorialization of a design tool workflow. The \emph{``Understandable and learnable''} category includes the need for unveiling complexity only when the designer is ready for it. Like players who become experts, however, designers do not want to be limited to defaults or to a high-level oversight role forever, and they eventually need access to full control. In several cases, Gold refers to the designers using the tool as ``players'' themselves. Thus, the designers argue that representing and thinking of AI in terms of characterization could aid designer intuition while creating the behavior, just as clear characterization can aid player understanding of the AI (as captured in \emph{``Communicate reasoning to the player''}).

\subsection{Implications for Computational Co-Creativity in Game Design}

This study reveals important implications for computational co-creativity tools for game design, and specifically for the design of AI characters. For instance, researchers are working towards transparency and explainability in AI~\cite{abdul_trends_2018}. However, these results support the need for going beyond transparency~\cite{amershi_guidelines_2019,shneiderman_human-centered_2020}, especially for computationally co-creative tools: they need to produce reliable results, as described by the \emph{``Reliable and efficient iteration''} category. Players can act unpredictably, and many machine learning algorithms might react unpredictably in response. The \emph{``Maintain an illusion of intelligence''} subcategory further elaborates on this, by noting game designers' need for fallback and contingency behaviors that can continue to preserve that illusion in such unexpected situations.

This is further reinforced by the \emph{``Designer has full control''} category, and particularly by its \emph{``Changes are preserved''} subcategory. It is not sufficient for the tool to produce consistent and reliable results: it should also provide designers with detailed control over the results in an intuitive language or interface. It may help for such tools to work with familiar game AI technologies, such as behavior trees and HFSMs, and to produce results that are modifiable and usable in existing game engines. This is also suggested by the \emph{``Enable standard tasks and workflows''} category.

The necessity for player experience as the primary final metric of success for AI behavior, as captured by the \emph{``Player experience determines success''} category, requires that the tools not merely use human data to train but also work with people to test their results. This means that many common measures of success, such as winning or ``perfect'' play, used in supervised machine learning or reinforcement learning are not sufficient to evaluate generated game AI. Researchers must enable designers to define ``success as being sometimes failing,'' and should also enable human-in-the-loop measures of AI behavior. This necessitates learning from a smaller number of examples and helping designers to create meaningful proxies for player behavior as success metrics or fitness functions. Just as players must learn a new game system and how to set goals in it, so too must game designers be assisted in defining goals in AI-enabled tools.

This has ethical implications, in part because many machine learning methods currently rely on large quantities of uncompensated or undercompensated human subjects data~\cite{sloane_participation_2020}. If player data is to be used as a success metric, it must be collected only with consent, and those who provide it must be compensated appropriately, protected from potential harms, and selected with consideration of under or over-representation, or over-burden, of particular groups~\cite{fairfield_big_2014}. The workers or players who provide such data should be empowered to retain their privacy, to provide feedback, and to be partners in the design process~\cite{sloane_participation_2020}. This presents both technological and regulatory challenges and requirements.

In general, these results highlight the need for intuitive, in-depth, and multi-way communication between the tool, the designer, and the player. Designers must be able to communicate AI behavior effectively to a player, but also the player must be able to communicate their experiences back to the designer and the tool. This multi-directional and multi-format communication presents a new challenge and opportunity for deeper engagement between tool, designer, and player in future computational co-creativity tool designs and methods.

\subsection{Limitations}
\label{sec:limitations}

We recruited two participants, which is on the small end of viable sample sizes for this type of qualitative study and thus limits the generalizability of these results. The designers consistently expressed agreement throughout the study, and their perspectives are informed by their work with other designers in games industry contexts, so it is likely that their opinions are relevant to designers with similar industry backgrounds and goals. Future work, however, will be necessary to confirm these requirements with larger populations of game designers and to test whether the requirements and design generated from this study will meet the needs of designers in somewhat different contexts.

For this study, we chose to work with game designers who had prior experience creating game AI as part of large, multi-disciplinary teams with significant resources. The researcher primarily responsible for participant interaction and qualitative analysis has worked as an engineer in similar industry contexts. We expected designers working in this privileged industry context to be familiar with supportive tools and engineering teams, and to know where they might still be improved. However, marginalized designers, and those with fewer resources to work with AI engineers or devote to AI development, will also have distinct goals, needs, and preferences that should be studied and taken into account when building tools to support them. Thus, this study does not represent the needs of marginalized developers in games, and future work should highlight and study their perspectives directly, with their participation and partnership.

Our contextualization of the study goals in terms of behavior trees and evolutionary algorithms influenced some of the designers' ideas. Moreover, our designers chose to focus on stealth games for their scenario, and their comments were sometimes specific to that genre -- other genres might reveal some different requirements. This may limit the generalizability of these results for other AI architectures and game types, and future studies should explore how designers would imagine co-creativity tools for those contexts. Finally, the necessity of performing the study remotely did cause some limitations in the designers' ability to collaborate in the participatory design process. It was more difficult for them to provide visual communication of their design ideas, though we did provide access to a virtual whiteboard and collaborative document editing software.

\section{Related Work}

This study and its results evince similarities to, and complications of, heuristics and requirements analyses for game design tools, and especially for computationally co-creative or mixed-initiative tools for designers. Confirmatory studies are necessary to validate and contextualize them, but several parallels are already evident. It is not clear whether the participants are aware of such parallels -- it is possible that our participants may be referencing existing heuristics for game design that they are familiar with, either through formal study or informal cultural knowledge.

Existing heuristics for game design emphasize the importance of feedback, starting with Malone's~\cite{malone_heuristics_1982} ``Provide performance feedback.'' The multiple heuristics developed for playability by Pinelle et al. \cite{pinelle_heuristic_2008,pinelle_usability_2009} and Desurvire et al. \cite{desurvire_game_2009,desurvire_user_2015,desurvire_using_2004} all emphasize the need for immediate, clear, and consistent feedback. Similarly, our participants emphasize the importance of feedback to communicate AI behavior to the player (the \emph{``Communicate reasoning to the player''} category), and also to communicate the tool's reasoning to the designer. While the latter can be considered feedback, it also relates to visibility of system status, as defined by Nielsen \cite{nielsen_ten_2005}. Also related to prior gameplay heuristics are the categories \emph{``Create varied, characterful behavior''} and \emph{``Empower game-specific AI behavior.''} These resonate with heuristic categories such as ``consistency in the world'' \cite{desurvire_game_2009}, ``in context'' \cite{desurvire_user_2015}, and ``immersion'' \cite{tondello_heuristic_2016}, all of which promote a game-specific, characterful, and narrative driven experience. They also relate to Suchman's~\cite{suchman_plans_1987} ``situatedness,'' which is highlighted for co-creative tools by the ``enactive'' model proposed by Davis et al.~\cite{davis_enactive_2015}. This indicates a connection wherein heuristics for a creative tool (a game AI design system) must facilitate fulfilling the heuristics for the creation (a game AI).

\emph{``Understandable and learnable''} and \emph{``Enable standard tasks and workflows''} bear relation to existing heuristics that tools should be consistent with standards so that users can intuitively apply preexisting knowledge or understanding \cite{nielsen_ten_2005,jimenez_usability_2016}. Another connection can be drawn between our \emph{``Help find and fix specific problems''} and the heuristic ``Help users recognize, diagnose, and recover from errors'' \cite{nielsen_ten_2005}. Both state that the system should help users identify their mistakes and recover from them. On the other hand, existing heuristics emphasize a need for error prevention \cite{nielsen_ten_2005} that appears to be less important to our participants. Specifically, the designers emphasize that good AI agents for human interaction do not exhibit ``perfect'' decision-making, encapsulated by \emph{``Player experience determines success,''} and Blue mentioned how they might find a surprisingly good behavior of an AI design by accident. This difference may reveal a point of departure between tools for AI design and tools in other domains. This echoes a point made in previous work \cite{jimenez_usability_2016} that different application domains require specific heuristics that apply to the needs of the users within the domain.

We also see several parallels between our categories and previously-proposed heuristics and models for co-creativity and mixed-initiative tools. Horvitz~\cite{horvitz_principles_1999} proposed that these tools should provide efficient means to directly change, undo, or override their results; our \emph{``Designer has full control''} category captures this insight. This is echoed in recent work from Lai, Latham, and Leymarie~\cite{lai_towards_2020} (``respect designer control''), and Amershi et al.~\cite{amershi_guidelines_2019} (``Support efficient correction.''). Several heuristics suggest the importance of efficient iteration and reliable results~\cite{horvitz_principles_1999,nielsen_ten_2005}, as does our \emph{``Reliable and efficient iteration''} category. They also often emphasize clearly communicating design details and feedback for designer actions, as our \emph{``Communicate reasoning to the designer''} category does; see, for instance, Amershi et al.'s~\cite{amershi_guidelines_2019} ``Make clear why the system did what it did'' and ``Convey the consequences of user actions,'' and Tondello et al.'s~\cite{tondello_heuristic_2016} ``Clear and Immediate Feedback''. Together, these promote controllability, in addition to transparency and explainability, as critical aspects of good computational co-creativity tools.

Several researchers also argue for controlling the level of detail and scaffolding learning, as captured by our \emph{``Understandable and learnable''} category and its \emph{``Complexity only as necessary''} sub-category. This starts from Malone's~\cite{malone_heuristics_1982} ``Optimal level of informational complexity'' and extends to Gentner and Grudin~\cite{gentner_design_1996}, and to Resnick et al.'s~\cite{resnick_design_2005} recommendations that tools provide a ``low threshold'' to help novice users get started, and be ``self-revealing'' of more sophisticated capabilities over time. most recently, Zhu et al.~\cite{zhu_player-ai_2021} recommend promotion of discovery-based learning. Our designers value this exploration-based revelation of detail; in session 3, for instance, Gold describes exploring the details underlying default agent behavior, saying ``And then if you get curious and you want to see at a lower level, you could keep expanding that, and seeing like, `Oh, that means that it set all of these variables to these defaults, and it activated these components, and...' you know under senses, like, `Oh, hearing! Ok, that –- that started off checked.'''

Finally, the mixed-initiative and generative design literature notes that one strength of such tools is in providing varied options for people to select between and modify~\cite{guzdial_friend_2019,craveirinha_exploring_2016,kreminski_generators_2019}. Kim and Foley~\cite{kim_providing_1993} argue that ``seeing many alternatives simultaneously, with metrics to compare them, supports decisions.'' Our participants surfaced this requirement as well, detailing cases where they want the system to ask questions, provide templates and options based on them, and enable characterful, varied design. Going beyond basic variety, the participants emphasized the idea of communicating through human-understandable ``character traits,'' as well as other means of comparing behavior. These requirements are captured in our \emph{``Anticipate and ask about needs''} and \emph{``Create varied, characterful behavior''} categories.

\section{Conclusion}

In this study we have developed, through a participatory design process of in-depth conversations with and between two game designers, a set of requirements and analyses for computational co-creativity tools, specifically focused on game AI design. Our detailed qualitative analysis of the results categorizes them into 13 categories, each with 2-3 sub-categories. We have explored the connections within, complications in, and implications of this categorization. We have also evinced connections to prior work in models and heuristics for game design, computational co-creativity, and mixed-initiative design tools, finding striking similarities and some complications of prior recommendations.

In future work, we plan to further develop and respond to these requirements for computational co-creativity tools. We plan to develop a tool for game AI design using these insights and categories as tentative heuristics, as well as the specific design elements proposed by the designers in this participatory design study, and to test its efficacy and usability. We believe that this tool should work within popular game engines, communicate through and build on common game AI design methods, and integrate methodology from mixed-initiative design tools.

The categorization presented and discussed above provides novel insights into the deep connections between game AI design and computational co-creativity tool requirements for game designers. Future work can further codify these design requirements into heuristics, specific design patterns, and novel, practical co-creativity tools. Finally, our analysis challenges designers and researchers in this area to think beyond current conceptions of AI transparency and ethics to call for additional guarantees of reliability and control, inspectability, and deep communication between and among the tool, the user, and the player or audience. To perform this research in a way that empowers people over corporate and AI systems, and especially those currently marginalized in AI and game development, we call for a focus on ethical and critical research in technological methods, societal change, and regulation to meet these new challenges.

\begin{acks}
Special thanks to Muhammad Ali at Northeastern for his contributions to a pilot study that helped inform this work.
\end{acks}

\bibliographystyle{ACM-Reference-Format}
\bibliography{references}

\end{document}